\documentclass[a4paper,cite]{article}
\usepackage{latexsym}

\usepackage[latin1]{inputenc}

\usepackage{graphicx}
\usepackage{color,pst-plot}

\usepackage{amsmath}
\usepackage {amsfonts,amssymb,amstext}

\newcommand{\lbl}[1]{\label{eq:#1}}
\newcommand{ \rf}[1]{(\ref{eq:#1})}

\newcommand{\be}{\begin{equation}}
\newcommand{\ee}{\end{equation}}{
\newcommand{\bea}{\begin{eqnarray}}
\newcommand{\eea}{\end{eqnarray}}
\newcommand{\setl}{\setlength\arraycolsep{2pt}}

\newcommand{\noi}{\noindent}
\newcommand{\nn}{\nonumber}
\newcommand{\ra}{\rightarrow}
\newcommand{\Ra}{\Rightarrow}

\newcommand{\cA}{{\cal A}}
\newcommand{\cB}{{\cal B}}

\newcommand{\cF}{{\cal F}}

\newcommand{\cM}{{\cal M}}

\newcommand{\cO}{{\cal O}}

\newcommand{\Ree}{\mbox{\rm Re}}

\newcommand{\hyper}[5]{\;_{#1}{\rm F}_{#2} \left(\left.\begin{matrix} {#3} \\ {#4} \end{matrix}\right| #5\right) }
\DeclareMathAlphabet{\eusm}{U}{}{}{}
\SetMathAlphabet\eusm{normal}{U}{eus}{m}{n}
\SetMathAlphabet\eusm{bold}{U}{eus}{b}{n}
\DeclareMathAlphabet{\mathpzc}{OT1}{pzc}{m}{it}

\input epsf
\setcounter{section}{0}

\setcounter{equation}{0}
\def\theequation{\arabic{section}.\arabic{equation}}

\begin{document}

\begin{titlepage}

\begin{flushright}
%\today  \\
CPT-2005/P020
\end{flushright}

\vspace*{0.2cm}
\begin{center}
{\Large {\bf Asymptotics of Feynman Diagrams \\ {\large and}\\ [0.2 cm] The Mellin--Barnes Representation
}}\\[2 cm]

{\bf Samuel Friot}~$^{ab}$, {\bf David Greynat}~$^a$ {\bf and  Eduardo de Rafael}~$^{a,\ c}$\\[1cm]

$^a$  {\it Centre  de Physique Th{\'e}orique~\footnote{Unit{\'e} Mixte de Recherche (UMR 6207) du CNRS et des Universit{\'e}s Aix Marseille 1, Aix Marseille 2 et sud Toulon-Var, affili{\'e}e {\`a} la FRUMAM.}\\
       CNRS-Luminy, Case 907\\
    F-13288 Marseille Cedex 9, France}\\[0.5cm]
$^b$ {\it Laboratoire de Math{\'e}matiques d'Orsay Universit{\'e} Paris-Sud~\footnote{Unit{\'e} Mixte de Recherche (UMR 8628) du CNRS.} \\Bât. 425 F-91405 Orsay
Cedex, France }\\[0.5cm]
$^c$ {\it Grup de F{\'\i}sica Te{\`o}rica and IFAE\\ Universitat Aut\`onoma de Barcelona\\ 08193 Barcelona, Spain \\

Instituci\'o Catalana de Recerca i Estudis Avan\c{c}ats (ICREA)}

\end{center}

\vspace*{3.0cm}

\begin{abstract}
It is shown that the integral representation of Feynman diagrams in terms of the traditional Feynman parameters, when combined with properties of the Mellin--Barnes representation and the so called {\it converse mapping theorem}, provide a very simple and efficient way to obtain the analytic asymptotic behaviours in both the large and small ratios of mass scales.
\end{abstract}

\end{titlepage}

%%%%%%%%%%%%%%%%%%%%%%%%%%%%%%%%%%
\section{\normalsize Introduction}\lbl{int}

The problem of extracting the asymptotic behaviour of Feynman diagrams, when ratios of mass scales in a diagram become large or small is an old one in Quantum Field Theory. 
Since the early days of Quantum Electrodynamics (QED), one has been confronted with it in almost every practical calculation of phenomenological interest. Although the discovery of dimensional regularization, together with the conceptual development of Effective Quantum Field Theory, have triggered several systematic technical approaches to this problem, the practical evaluation of even a few  asymptotic terms remains still a rather painful task. The purpose of this article  is to present a new approach to this problem, which appears to be astonishingly simple and provides a new physical insight.

The starting point is a systematic use of the old Feynman identity
\begin{equation}
	\frac{1}{ab}=\int_0^1 d\alpha \frac{1}{[\alpha a+(1-\alpha)b]^2}\,,
\end{equation}
which allows one to combine the products of all the  propagators $\frac{1}{a}$, $\frac{1}{b}$, ... in a Feynman graph into a single denominator where the internal circulating momenta $k_l$ appear in a quadratic form. The $k_l$--loop momenta integrals can then be done (if necessary in dimension $D=4-\epsilon$) by an appropriate shift of the loop momenta, which eliminates the terms of first degree in $k_l$, and then using standard diagonalization methods~(see e.g. ref.~\cite{ELOP66} for details). One can thus reduce a Feynman diagram (or set of diagrams) to the evaluation of a few dimensionless integrals over a fixed number $F$ of Feynman parameters of the type
\begin{equation}
	\cF (\rho_j)=\int_0^1 d\alpha_1 \int_0^1 d\alpha_2 \cdots\int_0^1 d\alpha_F \frac{N(\alpha_i)}{\left[\sum_j D_{j}(\alpha_i)\rho_j\right]^{n+ (p/q)\epsilon}}\,,
\end{equation}
where the $\rho_j$ denote scalar products of the external momenta and squared masses normalized to a fixed mass scale in the diagram, so that $\rho_0 =1$, and $n+(p/q)\epsilon >0$ with $n$, $p$ and $q$ positive integers. The difficulties, usually, start  at this point.

In order to illustrate the basic problem, let us consider the simplest case where there is only one $\rho$--parameter
\begin{equation}
	\sum_j D_{j}(\alpha_i)\rho_j=D_{0}(\alpha_i)+D_{1}(\alpha_i)\rho\,,
\end{equation}
and we want the behaviour of the integral $\cF (\rho)$ for $\rho\ll 1$. The root of the problem lies in the fact that the naive Taylor expansion (for $n=1,2,\dots$), 
\begin{equation}
\left(1+\frac{D_1(\alpha_i)}{D_0(\alpha_i)}\rho\right)^{-n-(p/q)\epsilon} =1-\left[n+(p/q)\epsilon\right]\frac{D_1(\alpha_i)}{D_0(\alpha_i)}\rho+\cO(\rho^2)\,,
\end{equation}
is useless because, practically always, the successive terms in these expansions lead to divergent integrals in the Feynman parameters $\alpha_i$. It is largely due to this fact that various techniques, other than the use of Feynman parameters, have been developed~\footnote{For recent reviews, see e.g. refs.~\cite{Smir01,Smir04} where other references can also be found.}. 

The basic feature of the approach that we shall present here is that it provides a regularization of the naive Taylor expansion in the $\rho$--parameter and at the same time it treats exactly the $\epsilon$--dependence of each coefficient in the $\rho$--expansion. The regulator is provided by  viewing $\cF (\rho)$  as an appropriate inverse Mellin transform associated to the simple analytic dependence that the Feynman parametric integrand provides on the $\rho$--parameter. As we shall see, the asymptotic behaviours of $\cF (\rho)$, both, for $\rho\ll 1$ and $\rho\gg 1$, are then encoded in the so called {\it converse mapping theorem}~\cite{FGD95} which establishes a precise relation between the singularities in the Mellin $s$--plane and the asymptotic behaviour(s) one is looking for.

The paper is organized as follows. In the next section we explain the basic features of the new approach that we propose as well as the {\it converse mapping theorem}. Section III illustrates this technique with the simple example of the lowest order vacuum polarization contribution from a light lepton (e) or a heavy lepton ($\tau$) to the muon anomaly. Section IV discusses the calculation of a typical {\it master integral} which often appears in Quantum Field Theory calculations. These examples offer the possibility of comparing our approach with other methods (much more tedious) which have been used previously in the literature to obtain asymptotic expansions of Feynman integrals. Finally,  Section V is dedicated to a three--loop calculation involving two ratios of mass scales, where an alternative type of a Mellin--Barnes representation is used.

%%%%%%%%%%%%%%%%%%%%%%%%%%%%%%%%%%%%%%%%%%%%%%%%%%%%%%%%%%%%%%%%%%%%%%%%
%%%%%%%%%%%%%%%%%%%%%%%%%%%%%%%%%%%%%%%%%%%%%%%%%%%%%%%%%%%%%%%%%%%%%%%%
\section{\normalsize The Feynman--Mellin--Barnes Representation}\lbl{FMB}
\setcounter{equation}{0}
\def\theequation{\arabic{section}.\arabic{equation}}

Let us pursue the study of the generic integral
\begin{equation}\lbl{feypar}
	\cF (\rho)=\int_0^1 d\alpha_1 \int_0^1 d\alpha_2 \cdots\int_0^1 d\alpha_F \frac{N(\alpha_i)}{\left[D_{0}(\alpha_i)+D_{1}(\alpha_i)\rho\right]^{\nu}}\,,\quad \nu=n+ (p/q)\epsilon\,,
\end{equation}
with only one mass--ratio $\rho$; i.e. the integral which we were left with in the introduction~\footnote{Although we shall later discuss a particular example with two mass ratios, we postpone the study of the general case with multiple mass scales to a forthcoming publication.} . We propose to apply the Mellin--Barnes representation~\cite{WW65}
\begin{equation}\lbl{MBR}
	\frac{1}{\left(1+X\right)^{\nu}}=\frac{1}{2\pi i}\int\limits_{c-i\infty}^{c+i\infty} ds\  \left(X\right)^{-s}\ \frac{\Gamma(s)\Gamma(\nu-s)}{\Gamma(\nu)}\,,
\end{equation}
with the integration path along the imaginary axis
defined in the {\it fundamental strip}~: $c=\text{Re}(s)\;\in\;]0,\nu[$,
to the integrand in Eq.~\rf{feypar}. This results in a new integral representation
\begin{equation}\lbl{feyparMB}
	\cF (\rho)=\frac{1}{2\pi i}\int_0^1 d\alpha_1 \int_0^1 d\alpha_2 \cdots\int_0^1 d\alpha_F \ \frac{N(\alpha_i)}{\left[D_{0}(\alpha_i)\right]^{\nu}}\int\limits_{c-i\infty}^{c+i\infty} ds\  \rho^{-s} \left(\frac{D_{0}(\alpha_i)}{D_1(\alpha_i)}\right)^{s}\ \frac{\Gamma(s)\Gamma(\nu-s)}{\Gamma(\nu)}\,.
\end{equation}
The crucial property of this representation is that it decouples the dependence on the $\rho$--parameter from the one on the Feynman parameters $\alpha_i$. The integrals over the Feynman parameters $\alpha_i$ are now those of a massless theory, and they are regularized by the $s$--dependence of the integrand (much the same as dimensional regularization does with integrals over the loop momenta). In fact, as we shall  see later in a few examples, these integrals are often rather simple and in many cases they result in products of $\Gamma$--functions. 

Altogether, we obtain an integral representation
\begin{equation}
	\cF (\rho)=\frac{1}{2\pi i}\int\limits_{c-i\infty}^{c+i\infty} ds\  \rho^{-s}\ 	\cM[\cF](s)\,,
\end{equation}
where
\begin{equation}
\cM[\cF](s)=\int_0^1 d\alpha_1 \int_0^1 d\alpha_2 \cdots\int_0^1 d\alpha_F \ \frac{N(\alpha_i)}{\left[D_{0}(\alpha_i)\right]^{\nu}}\left(\frac{D_0(\alpha_i)}{D_1(\alpha_i)}\right)^{s}\frac{\Gamma(s)\Gamma(\nu-s)}{\Gamma(\nu)}
\end{equation}
is the Mellin transform of the Feynman integral at the start in Eq.~\rf{feypar}.
In full generality, the function $\cM[\cF](s)$ is a meromorphic function in the complex $s$--plane and, in perturbation theory, its singularities (poles and/or multiple poles) lie in the $\Ree(s)$--axis; outside of the {\it fundamental strip}~\footnote{In some cases, as we shall see later, the singular structure of the function $\cM[\cF](s)$ associated to a specific Feynman integral  may restrict the width of the {\it fundamental strip}. The case where no {\it fundamental strip} exists which, so far, we have not encountered in the Quantum Field Theory examples that we have analyzed,  has also been considered in the mathematical literature~\cite{BH75}.}. The position of the  poles on the r.h.s. of the {\it fundamental strip}, their multiplicity and their residues encode the asymptotic behaviour of $\cF (\rho)$ for $\rho\gg 1$; the poles on the l.h.s. and their residues, those of the asymptotic behaviour of $\cF (\rho)$ for $\rho\ll 1$. The precise form of the encoding~\cite{Doe55} is spelled out by a theorem which in the mathematical literature goes under the name of the {\it converse mapping theorem}~(see ref.~\cite{FGD95} for a proof of this theorem), and which we next discuss.    
   
%%%%%%%%%%%%%%%%%%%%%%%%%%%%%%%%%%
\subsection{\normalsize {\sc The Converse Mapping Theorem}}\lbl{conmap}

The theorem in question relates the singularities of the function  $\cM[\cF](s)$  in the complex $s$--plane to the 	asymptotic behaviour of the Feynman graph $\cF (\rho)$ as follows:

\begin{itemize}
	\item {\sc Right--Hand--Side Singularities $\Ra$ Expansion for $\rho\ra\infty$}
	
With   $\xi(\nu)\in\mathbb{R}$ and $k\in\mathbb{N}$, the function $\cM[\cF](s)$ in the r.h.s. of the {\it fundamental strip} has  a singular expansion~\footnote{ The singular expansion (or singular series) of a meromorphic function is a formal series collecting the singular elements
at all poles of the function (a singular element being the truncated
Laurent's series (at $\mathcal{O}(1)$) of the function at a given pole)
and it is denoted by the symbol $\asymp$~\cite{FGD95}.} of the type (ordered in increasing values of $\xi$):
\be
\cM[\cF](s)\asymp\sum_{\xi}\sum_{k}\frac{	\mathsf{a}_{\xi,k}}{(s-\xi)^{k+1}}\,.
\ee
The corresponding asymptotic behaviour of $\cF (\rho)$ for $\rho$ large (ordered in increasing powers of $\xi$) is then:
\begin{equation}
	\cF (\rho)\underset{{\rho\ra\infty}}{\thicksim}\sum_{\xi}\sum_{k} \frac{(-1)^{k+1}}{k!}\ \mathsf{a}_{\xi,k}\ \rho^{-\xi}\ \log^{k}\rho\,.
\end{equation}

	\item {\sc Left--Hand--Side Singularities $\Ra$ Expansion for $\rho\ra 0$}
	
With   $\xi(\nu)\in\mathbb{R}$ and $k\in\mathbb{N}$, the function $\cM[\cF](s)$ in the l.h.s. of the {\it fundamental strip} has  a singular expansion of the type (ordered in increasing values of $\xi$):
\be
\cM[\cF](s)\asymp\sum_{\xi}\sum_{k}\frac{	\mathsf{b}_{\xi,k}}{(s+\xi)^{k+1}}\,.
\ee
The corresponding asymptotic behaviour of $\cF (\rho)$ (ordered in increasing powers of $\xi$) is then:
\begin{equation}
	\cF (\rho)\underset{{\rho\ra 0}}{\thicksim}\sum_{\xi}\sum_{k} \frac{(-1)^{k}}{k!}\ \mathsf{b}_{\xi,k}\ \rho^{\xi}\ \log^{k}\rho\,.
\end{equation}
\end{itemize}

The basic steps which we propose are: i) the old--fashioned Feynman parameterization, with explicit integration over the loop momenta, and only then to use a conveniently chosen  Mellin--Barnes representation; ii) the factorized Feynman parametric integrals, which are then $s$--regularized and $\rho$--independent, appear to be remarkably simple; iii) the use of the {\it converse mapping theorem} which encodes the relation between the singular behaviour in the Mellin $s$--plane and the full asymptotic behaviours of the initial Feynman integral for small or large ratios of mass parameters~\footnote{To our knowledge, the use of the Mellin-Barnes representation in Quantum Field Theory was first proposed in refs.~\cite{BW63,TY63}. Most of the recent calculations in the literature which use this technique (see e.g. ref.~\cite{BD91} and refs.~\cite{Smir01,Smir04}) introduce a Mellin--Barnes representation for each free propagator in a diagram. Calculations in an approach similar to the one we propose here can be found e.g. in refs.~\cite{GHW96,AAGW04}. None of these references, however, uses the rigorous short--cut provided by the {\it converse mapping theorem}.}. These asymptotic series can be obtained without the explicit knowledge of the exact analytic expressions. It is also possible, if necessary, to extract an arbitrary term in those series  by looking at the residues of the appropriate pole (or multiple pole) at the right or at the left of the {\it fundamental strip}, without having to do the calculation of  the corresponding lower order terms.

The rest of this paper is dedicated to a few interesting examples which illustrate the simplicity of this approach; but there are many other possible applications one can think of which  we invite the reader to consider.

%%%%%%%%%%%%%%%%%%%%%%%%%%%%%%%%%%
\section{\normalsize  Vacuum Polarization Contributions to the Muon Anomaly}\lbl{VPM}

Historically, this is the first example where the problem of extracting an asymptotic behaviour in the ratio of two masses in Quantum Field Theory appeared. The relevant Feynman graphs are shown in Fig.~1, where the internal fermion $l$ in the vacuum polarization loop can either be an electron (large ratio $m_{\mu}/m_e$) or a $\tau$ (small ratio  $m_{\mu}/m_{\tau}$). Their contribution to the muon anomaly is known analytically~\cite{El66}~\footnote{The history of vacuum polarization contributions can be traced back from the review articles in refs.~\cite{LPdeR72, Pas05}}. Here we are only considering these contributions once more, as an illustration of the simplicity of our approach.

%%%%%%%%%%%%%%%%%%%%%%%%%%%%%%%%%%%%%%
\begin{figure}[h]

\begin{center}
\includegraphics[width=0.6\textwidth]{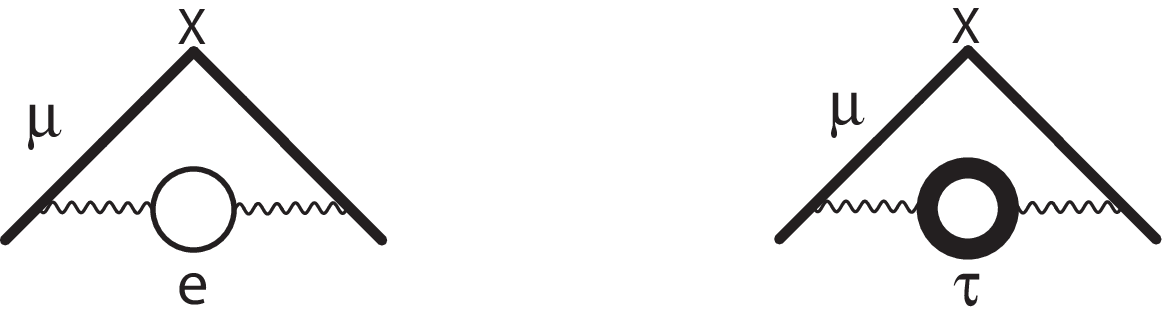}

\vspace*{0.5cm}
{\bf Fig.~1} {\it\small  Vacuum Polarization contributions to the Muon Anomaly\\ from a Small internal mass (electron) and from a Large internal mass (tau).
}
\end{center}

\end{figure}
%%%%%%%%%%%%%%%%%%%%%%%%%%%%%%%%%%%%%%

For the purpose we are concerned with here, it is convenient to start with the Feynman parameterization~\cite{LdeR69,LPdeR72} (which already takes into account the on--shell renormalization  of the vacuum polarization subgraph):
\begin{equation}\lbl{Fpr}
a_{\mu}^{\rm\tiny v.p.} = \left(\frac{\alpha}{\pi}\right)^2 \int_0^1 \frac{dx}{x}(1-x)(2-x)\int_0^1 dy y(1-y) \frac{1}{1+\frac{m_l^2}{m_{\mu}^2}\frac{1-x}{x^2}\frac{1}{y(1-y)}}\,.	
\end{equation}
This is a particular case of an integral like the one in Eq.~\rf{feypar} corresponding to the case of two Feynman parameters with $\alpha_{1}=x$,  $\alpha_{2}=y$, $\rho=\frac{m_l^2}{m_{\mu}^2}$ and $\nu=1$. 

The Mellin--Barnes representation in Eq.~\rf{MBR}, when applied to the last factor in the integrand in Eq.~\rf{Fpr}, gives
\begin{equation}
\frac{1}{1+\frac{m_l^2}{m_{\mu}^2}\frac{1-x}{x^2}\frac{1}{y(1-y)}}=\frac{1}{2\pi i}\int\limits_{c-i\infty}^{c+i\infty} ds \left(\frac{m_l^2}{m_{\mu}^2} \right)^{-s}\left[y(1-y)\frac{x^2}{1-x} \right]^{s} \Gamma(s)\Gamma(1-s)\,.
\end{equation}
The integrals over the Feynman parameters reduce now to simple Beta functions $\left[{\rm B}(u,v)=\frac{\Gamma(u)\Gamma(v)}{\Gamma(u+v)}\right]$.  
Altogether, we have the simple representation
\begin{equation}
	a_{\mu}^{\rm\tiny v.p.}=\frac{1}{2\pi i}\int\limits_{c-i\infty}^{c+i\infty} ds \left(\frac{m_l^2}{m_{\mu}^2} \right)^{-s}\cM[a_{\mu}^{\rm\tiny v.p.}](s)\,,
\end{equation}
with
\begin{equation}\lbl{singsvp}
\cM[a_{\mu}^{\rm\tiny v.p.}](s)=\frac{ (1-s)\left[\Gamma(s)\Gamma(1-s) \right]^2}{(2+s)(1+2s)(3+2s)}\,,
\end{equation}
the function which encodes the asymptotic behaviours we are looking for.

%%%%%%%%%%%%%%%%%%%%%%%%%%%%%%%%%%%%%%
\begin{figure}[h]

\begin{center}
\rotatebox{-90}{\includegraphics[width=0.35\textwidth]{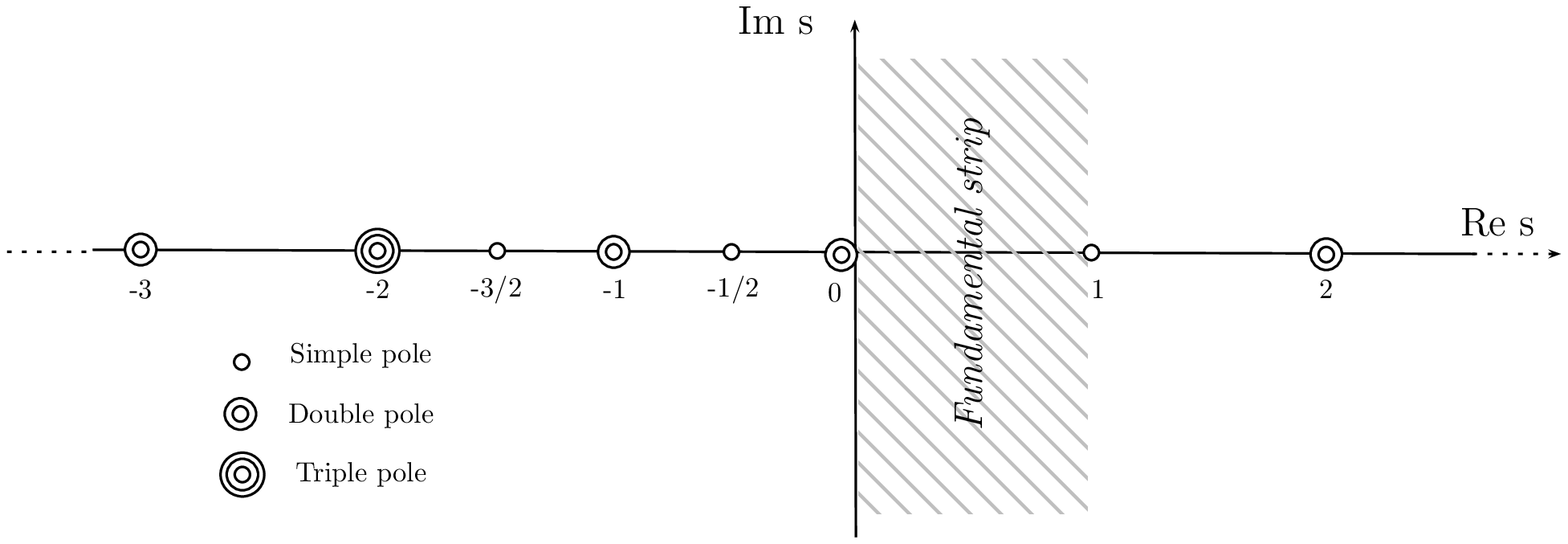}}

\vspace*{0.5cm}
{\bf Fig.~2} {\it\small  Singularities of the function $\cM[a_{\mu}^{\rm\tiny v.p.}](s)$ (see Eq.~\rf{singsvp}) in the complex $s$--plane.
}
\end{center}

\end{figure}
%%%%%%%%%%%%%%%%%%%%%%%%%%%%%%%%%%%%%%

As illustrated in Fig.~2, the singularities of $\cM[a_{\mu}^{\rm\tiny v.p.}](s)$ in the r.h.s. of the {\it fundamental strip}: $c=\text{Re}(s)\;\in\;]0,1[$, appear as a pole at $s=1$ and as double and single poles at $s=n$ for $n=2,3,\dots\,.$ The nearer singularities to the {\it fundamental strip}, from the right, fix the leading terms in the asymptotic expansion when $m_l\gg m_{\mu} $ and they can be obtained by simple inspection of Eq.~\rf{singsvp}. We cannot refrain, however, from giving the explicit form of the full singular series, which can be obtained rather simply:
\begin{equation}\lbl{expsing}
\cM[a_{\mu}^{\rm\tiny v.p.}](s)\!\asymp\!\sum_{n=0}^\infty\!\left[\!\frac{1}{s\!-\!(n+1)}\frac{-45\!+\!28n^2\!+\!8n^3}{[(3\!+\!n)(3\!+\!2n)(5\!+\!2n)]^2}\!+\!\frac{1}{[s\!-\!(n+1)]^2}\frac{(-n)}{(3\!+\!n)(3\!+\!2n)(5\!+\!2n)}\!\right]\,.
\end{equation}
According to the {\it converse mapping theorem} this singular series encodes the asymptotic expansion~\footnote{The  leading term of this expansion can be physically understood~\cite{LdeR68} as the convolution of the slope of the vacuum polarization at the origin with the contribution to the muon anomaly from a heavy--photon.}:
\begin{equation}\lbl{vpgg}
a_{\mu}^{\rm\tiny v.p.}\!\!\underset{m_\tau \gg m_{\mu}}{\thicksim}\!\!\left(\frac{\alpha}{\pi}\right)^2\sum_{n=0}^\infty	\left(\frac{m_\mu^2}{m_\tau^2}\right)^{\! n+1}\!\!\left[\frac{45\!-\!28n^2\!-\!8n^3}{[(3+n)(3\!+\!2n)(5\!+\!2n)]^2}\!+\!\frac{n}{(3\!+\!n)(3\!+\!2n)(5\!+\!2n)}\ln\frac{m_\tau^2}{m_\mu^2}\!\right]\,;
\end{equation}
a result which allows for an immediate evaluation of as many terms as one wishes in the asymptotic expansion.

The singularities of $\cM[a_{\mu}^{\rm\tiny v.p.}](s)$ at the l.h.s. of the {\it fundamental strip}~: $c=\text{Re}(s)\;\in\;]0,1[$, appear as single poles at $s=-1/2$ and $s=-3/2$; a triple, double and single pole at $s=-2$;  and double and single poles at $s=0$, $s=-1$ and $s=-n$ for $n=3,4,\dots\,.$Again, it is rather easy to obtain the corresponding singular series:

{\setl
\bea
\cM[a_{\mu}^{\rm\tiny v.p.}](s) & \asymp & \frac{\pi^2}{4} \frac{1}{s+\frac{1}{2}} -\frac{5\pi^2}{4} \frac{1}{s+\frac{3}{2}}+\frac{1}{(s+2)^3}+ \frac{7}{3}\frac{1}{(s+2)^2}+ \left(\frac{44}{9} + \frac{\pi^2}{3}\right) \frac{1}{s+2}  \nonumber\\
&  & \hspace*{-1cm}+\sum_{\substack{n=0 \\ n\neq 2}}^{\infty}\!\left[\! \frac{1+n}{(2-n)(1-2n)(3-2n)}\frac{1}{(s+n)^2} \!+\! \frac{-25+32n+4n^2-8n^3}{(2-n)^2(1-2n)^2(3-2n)^2}\frac{1}{s+n}\!\right]\!\,,
\eea}

\noi
which according to the {\it converse mapping theorem} encodes the asymptotic expansion~\footnote{The leading $\log$--term  in this expansion is in fact fixed by simple renormalization group arguments~\cite{LdeR74}.}

{\setl
\bea\lbl{vpll}
	a_{\mu}^{\rm\tiny v.p.} &\underset{m_e \ll m_{\mu}}{\thicksim} & \left(\frac{\alpha}{\pi}\right)^2\left\{\frac{1}{6}\ln\frac{m_{\mu}^2}{m_e^2}-\frac{25}{36}+\frac{\pi^2}{4} \frac{m_e}{m_\mu}+\left(\frac{m_e^2}{m_{\mu}^2}\right)\left[-2\ln\frac{m_{\mu}^2}{m_e^2}+3 \right]-\frac{5\pi^2}{4}\left(\frac{m_e}{m_\mu}\right)^3 \right. \nn \\ & &  \!+ \left(\frac{m_e^2}{m_\mu^2}\right)^2\left[\frac{1}{2}\ln^2\frac{m_e^2}{m_\mu^2}- \frac{7}{3}\ln\frac{m_e^2}{m_\mu^2}+ \frac{44}{9} + \frac{\pi^2}{3}\right]\nn \\
& & \left. \!+\!\!\sum_{n=3}^{\infty}\!\left(\frac{m_e^2}{m_\mu^2}\right)^{\!\!n}\!\!\left[ \frac{1+n}{(2-n)(1-2n)(3-2n)}\ln\!\frac{m_\mu^2}{m_e^2}\! +\! \frac{-25+32n+4n^2-8n^3}{(2-n)^2(1-2n)^2(3-2n)^2}\!\right]\!\right\}\!\!\,.
\eea}

The results in Eqs.~\rf{vpgg} and \rf{vpll} agree with the asymptotic expansions given in ref.~\cite{LMS93}, obtained from the exact analytic expression in~\cite{El66}. The new feature which we have shown here, with this well known example, is how a full asymptotic series can be obtained, \underline{directly} and in a \underline{simple} way, without knowledge of the exact analytic expression. 
The reader can judge the simplicity of our method, by comparing it to other ways that a few terms of the same asymptotic expansions have been obtained in the literature (see e.g. ref~\cite{CS98}, where the so--called {\it method of regions}~\cite{BS98} is used). 

%%%%%%%%%%%%%%%%%%%%%%%%%%%%%%%%%%%%%%%%%%%%%%%%%%%%%%%%%%%%%%%%%%
%%%%%%%%%%%%%%%%%%%%%%%%%%%%%%%%%%%%%%%%%%%%%%%%%%%%%%%%%%%%%%%%%%
\section{\normalsize  Calculation of a Master Integral}\lbl{MasIn} 
\setcounter{equation}{0}
\def\theequation{\arabic{section}.\arabic{equation}}

The next example we wish to discuss concerns the integral
\begin{equation}\lbl{master}
J(p^2,m^2,t)=\frac{1}{(2\pi)^{2D}}\int d^Dk_1\int d^Dk_2 \frac{1}{\left(k_1^2-t\right)\left[ (k_1-k_2)^2 -m^2 \right](k_2-p)^2}\;,
\end{equation}
in dimensional regularization $D=4-\epsilon$. This is a typical {\it master
 integral} which appears in many two loop calculations as a result of a systematic use of recurrence relations obtained from integrations by parts in $D$--dimensions~\footnote{Examples of such calculations can be found e.g. in refs.~\cite{BFT93,CCLR98} and references therein.}. Here we shall be concerned with the case where $p^2=m^2$ and the dependence of the integral on the massless ratio $\rho=\frac{m^2}{t}$ for $\rho\ll 1$. In fact, it is precisely the calculation of this integral which triggered our ideas on the approach that we are advocating here.

A standard Feynman parameterization of the three propagators in Eq.~\rf{master} results, after integration over the loop momenta $k_1$ and $k_2$, in a two dimensional Feynman integral which, a priori, looks rather complicated. We split it into two terms, so as to reduce it to integrals of the generic type in Eq.~\rf{feypar}:
\begin{equation}\lbl{JMMT}
	J(m^2,m^2,t)=\left( \frac{i}{16\pi^2}\right)^2 \left(4\pi\right)^\epsilon \frac{\Gamma(\epsilon)}{1-\epsilon}\left(\frac{\mu^2}{t}\right)^\epsilon\;t\ \left[
	\cF^{(1)}\left(\frac{m^2}{t}\right)+\frac{m^2}{t}\  \cF^{(2)}\left(\frac{m^2}{t}\right) \right]\,,
\end{equation}
with $\left[  Y=1-y(1-y)\right]$
\be\lbl{J1}
\cF^{(1)}\left(\frac{m^2}{t}\right)=\int_0^1 dx \int_0^1 dy\ \frac{x^{-\frac{\epsilon}{2}}y^{1-\epsilon}\left( 1- x Y\right)^{-2 + \frac{\epsilon}{2}}}{\left[1 + \frac{m^2}{t}\frac{(1-y)^2}{y}\frac{1-x(1-y)}{1-x Y} \right]^{\epsilon}} \,,
\ee  
and
\be\lbl{J2}
\cF^{(2)}\!\left(\frac{m^2}{t}\right)=\int_0^1 dx \int_0^1 dy\ \frac{x^{-\frac{\epsilon}{2}}y^{-\epsilon}\left( 1-x Y\right)^{-3 + \frac{\epsilon}{2}}  (1-y)^2 [1-x(1-y)] }{\left[1 + \frac{m^2}{t} \frac{(1-y)^2}{y}\frac{1-x(1-y)}{1-x Y} \right]^{\epsilon}} \,.
\ee
 
Following the methodology explained in Section II, we apply the Mellin--Barnes representation in Eq.~\rf{MBR} to the denominator of the integrands in Eqs.~\rf{J1} and \rf{J2}:
\begin{align}
\label{MB}
\frac{1}{\left[1 \!+\! \frac{m^2}{t} \frac{(1\!-\!y)^2}{y}\frac{1\!-\!x(1\!-\!y)}{1-x Y} \right]^{\epsilon}}
\! =\!  \frac{1}{2\pi i}\!\int\limits_{c-i\infty}^{c+i\infty} \! \! ds \!\left(\frac{m^2}{t}\right)^{-s}\! \left[\frac{(1-y)^2}{y}\frac{1-x(1-y)}{1-x Y}\right]^{-s} \frac{\Gamma(s) \Gamma\left(\epsilon\!-\!s\right)}{\Gamma\left(\epsilon\right)}\!\,;
\end{align}
and next consider the two integrals $\cF^{(1)}\left(\frac{m^2}{t}\right)$ and $\cF^{(2)}\left(\frac{m^2}{t}\right)$ separately.

\begin{itemize}
	\item {\sc The Integral $\cF^{(1)}\left(\frac{m^2}{t}\right)$}
	
	The integral over the Feynman $x$--variable in $\cF^{(1)}\left(\frac{m^2}{t}\right)$ can be done in terms of a Gauss hypergeometric function which, for convergence purposes, keeping in mind the fact that we still have to integrate over the $y$--variable in the range $[0,1]$, we choose to express in the form~\cite{AK26}
\be
\int_0^1\! dxx^{-\frac{\epsilon}{2}}\left[1\!-\!x(1\!-\!y)\right]^{-s}\!\left(1\!-\!xY\right)^{-2+s+\frac{\epsilon}{2}}\!=\!\text{B}\left(1,1\!-\!\frac{\epsilon}{2}\right)\!  y^{\frac{\epsilon}{2}-1}
\hyper{2}{1}{1\!-\!\frac{\epsilon}{2},2\!-\!\frac{\epsilon}{2}\!-\!s}{2-\frac{\epsilon}{2}}{y}\!\,.
\ee
Expanding this hypergeometric function in powers of $y$, one can do the $y$--integration term by term. The resummation of the series leads then to a simple product of $\Gamma$--functions~\cite{AK26}.
The overall result is a simple Mellin--Barnes representation for the $\cF^{(1)}\left(\frac{m^2}{t}\right)$ integral
\be\lbl{BM1}
\cF^{(1)}\left(\frac{m^2}{t}\right)=
\frac{1}{2\pi i}\int\limits_{c-i\infty}^{c+i\infty} \! \! ds \;\left(\frac{m^2}{t}\right)^{-s} \; \cM[\cF^{(1)}](s) \,,
\ee
with $\cM[\cF^{(1)}](s)$  defined by the product of $\Gamma$--functions:
\be\lbl{BMR1}
\cM[\cF^{(1)}](s)=\frac{\Gamma\left(1-\frac{\epsilon}{2}\right)}{\Gamma(\epsilon)}
\frac{\Gamma(s) \Gamma\left(\epsilon-s\right)\Gamma(1-2s)\Gamma\left(1+s-\frac{\epsilon}{2}\right)\Gamma\left(\frac{\epsilon}{2}-s\right)}{\Gamma\left(2-s-\frac{\epsilon}{2}\right)\Gamma(1-s)}\,.
\ee
Notice that, because of the presence of the $\Gamma\left(\frac{\epsilon}{2}-s\right)$ factor, which appears as a result of the integration over the Feynman parameters, the {\it fundamental strip} in the Mellin Barnes representation in Eq.~\rf{BM1} is now restricted to $c=\text{Re}(s)\;\in\;]0,\frac{\epsilon}{2}[$.

	\item {\sc The Integral $\cF^{(2)}\left(\frac{m^2}{t}\right)$}
	
	A procedure entirely similar to the one just described for the function $\cF^{(1)}\left(\frac{m^2}{t}\right)$ leads to the corresponding representation
\be\lbl{BM2}
\cF^{(2)}\left(\frac{m^2}{t}\right)=
\frac{1}{2i\pi}\int\limits_{c-i\infty}^{c+i\infty} \! \! ds \;\left(\frac{m^2}{t}\right)^{-s} \; \cM[\cF^{(2)}](s)\,,
\ee
with $ \cM[\cF^{(2)}](s)$ defined by the product of $\Gamma$--functions:
\be\lbl{BMR2}
\cM[\cF^{(2)}](s)=\frac{\Gamma\left(1-\frac{\epsilon}{2}\right)}{\Gamma(\epsilon)}
\frac{\Gamma(s)\Gamma(\epsilon-s)\Gamma\left(s-\frac{\epsilon}{2}\right)\Gamma\left(3-2s\right)\Gamma\left(1+\frac{\epsilon}{2}-s\right)}{\Gamma\left(3-s-\frac{\epsilon}{2}\right)\Gamma\left(2-s\right)}\,,
\ee
and the {\it fundamental strip} in the Mellin--Barnes representation in Eq.~\rf{BM2} restricted to 
$c=\text{Re}(s)\;\in\;]\frac{\epsilon}{2},\epsilon[$.	
 
\end{itemize}

The singularities of $\cM[\cF^{(1)}](s)$ in Eq.~\rf{BMR1} in the l.h.s. of the {\it fundamental strip}~ $c=\text{Re}(s)\;\in\;]0,\frac{\epsilon}{2}[$, appear as a double series of simple poles at $s=-n$ and $s=-n-1+\frac{\epsilon}{2}$, $n\in\mathbb{N}$, with residues given by the singular expansion

{\setl
\bea
\cM[\cF^{(1)}](s) & \asymp & \frac{\Gamma\left(1-\frac{\epsilon}{2}\right)}{\Gamma(\epsilon)}
\left[\sum_{n=0}^\infty \frac{(-1)^n}{n!}\; \frac{ \Gamma\left(\epsilon+n\right)\Gamma(1+2n) \Gamma\left(1-n-\frac{\epsilon}{2}\right) \Gamma\left(\frac{\epsilon}{2}+n\right)}{\Gamma\left(2+n-\frac{\epsilon}{2}\right)\Gamma(1+n)}\;\frac{1}{s+n} \right.\nonumber\\
& &\!\!\!\!\!\! + \left.\sum_{n=0}^\infty (-1)^n\;\frac{\Gamma(-n-1+\frac{\epsilon}{2}) \Gamma\left(n+1+\frac{\epsilon}{2}\right)\Gamma(3+2n-\epsilon)}{\Gamma\left(3+n-\epsilon\right)\Gamma(2+n-\frac{\epsilon}{2})}\;\frac {1}{s+n+1-\frac{\epsilon}{2}} \right]\;.
\eea}

\noi
The {\it converse mapping theorem} associates to this singular series the asymptotic expansion

{\setl
\bea\lbl{result1}
\hspace*{-1cm}\cF^{(1)}\left(\frac{m^2}{t}\right)& \underset{m^2 \ll~t} {\thicksim} & 
\frac{\Gamma\left(1-\frac{\epsilon}{2}\right)}{\Gamma(\epsilon)} \left[\sum_{n=0}^\infty \frac{(-1)^n}{n!}\; \frac{ \Gamma\left(\epsilon+n\right)\Gamma(1+2n) \Gamma\left(1-n-\frac{\epsilon}{2}\right) \Gamma\left(\frac{\epsilon}{2}+n\right)}{\Gamma\left(2+n-\frac{\epsilon}{2}\right)\Gamma(1+n)}\;\left(\frac{m^2}{t}\right)^n\right. \nonumber\\
& & +\left.\sum_{n=0}^\infty (-1)^n\;\frac{\Gamma(-n-1+\frac{\epsilon}{2}) \Gamma\left(n+1+\frac{\epsilon}{2}\right)\Gamma(3+2n-\epsilon)}{\Gamma\left(3+n-\epsilon\right)\Gamma(2+n-\frac{\epsilon}{2})}\;\left(\frac{m^2}{t}\right)^{n+1-\frac{\epsilon}{2}} \right]\;.
\eea}

The singularities of $\cM[\cF^{(2)}](s)$ in Eq.~\rf{BMR2} in the l.h.s. of the {\it fundamental strip}~ $c=\text{Re}(s)\;\in\;]\frac{\epsilon}{2},\epsilon[$, appear as a double series of simple poles at $s=-n$ and $s=-n+\frac{\epsilon}{2}$, $n\in\mathbb{N}$, with residues given by the singular expansion

{\setl
\bea
\cM[\cF^{(2)}](s) & \asymp & 
\frac{\Gamma\left(1-\frac{\epsilon}{2}\right)}{\Gamma(\epsilon)}\left[\sum_{n=0}^\infty \frac{(-1)^n}{n!} \frac{\Gamma(\epsilon+n)\Gamma\left(-n-\frac{\epsilon}{2}\right)\Gamma\left(3+2n\right)\Gamma\left(1+\frac{\epsilon}{2}+n\right)}{\Gamma\left(3+n-\frac{\epsilon}{2}\right)\Gamma\left(2+n\right)} \frac{1}{s+n} \right.\nonumber\\
& & \left. + \sum_{n=0}^\infty (-1)^n\frac{\Gamma(-n+\frac{\epsilon}{2})\Gamma(\frac{\epsilon}{2}+n)\Gamma\left(3+2n-\epsilon\right)}{\Gamma\left(3+n-\epsilon\right)\Gamma\left(2+n-\frac{\epsilon}{2}\right)}\frac{1}{s+n-\frac{\epsilon}{2}} \right]\;,
\eea}

\noi
with which, the {\it converse mapping theorem} associates the asymptotic behaviour

{\setl
\bea\lbl{result2}
\hspace*{-0.7cm}\cF^{(2)}\left(\frac{m^2}{t}\right)&\underset{m^2 \ll~t}{{\thicksim}} & \frac{\Gamma\left(1-\frac{\epsilon}{2}\right)}{\Gamma(\epsilon)}\left[\sum_{n=0}^\infty \frac{(-1)^n}{n!} \frac{\Gamma(\epsilon+n)\Gamma\left(-n-\frac{\epsilon}{2}\right)\Gamma\left(3+2n\right)\Gamma\left(1+\frac{\epsilon}{2}+n\right)}{\Gamma\left(3+n-\frac{\epsilon}{2}\right)\Gamma\left(2+n\right)}\left(\frac{m^2}{t}\right)^{n} \right.\nonumber\\
& & \left. + \sum_{n=0}^\infty (-1)^n\frac{\Gamma(-n+\frac{\epsilon}{2})\Gamma(\frac{\epsilon}{2}+n)\Gamma\left(3+2n-\epsilon\right)}{\Gamma\left(3+n-\epsilon\right)\Gamma\left(2+n-\frac{\epsilon}{2}\right)}\left(\frac{m^2}{t}\right)^{n-\frac{\epsilon}{2}} \right]\;.
\eea}

\noi

The results in Eqs.~\rf{result1} and \rf{result2}, when incorporated in Eq.~\rf{JMMT} give the full asymptotic expansion of the master integral $J(m^2,m^2,t)$ for $\frac{m^2}{t}\ra 0$, with coefficients which have a full exact dependence in the $\epsilon$--variable. Again, the relative simplicity of this calculation using our approach, should be compared to the complicated expressions obtained by other methods in the literature~\cite{JK04}. We have checked that our results coincide with those  given in ref.~\cite{CCLR98} for the terms of $\cO(1/\epsilon^2)$ and $\cO(1/\epsilon)$, and in ref.~\cite{CGZ02} for the term of $\cO(\epsilon^0)$, when their corresponding expressions are expanded for $m^2 \ll t$.

%%%%%%%%%%%%%%%%%%%%%%%%%%%%%%%%%%%%%%%%%%%%%%%%%%%%%%%
%%%%%%%%%%%%%%%%%%%%%%%%%%%%%%%%%%%%%%%%%%%%%%%%%%%%%%%
\section{\normalsize  Three Loop Calculation}\lbl{3loop} 
\setcounter{equation}{0}
\def\theequation{\arabic{section}.\arabic{equation}}

In this last example, we shall consider the contribution to the muon anomaly
from the vacuum polarization correction to the photon propagator induced, simultaneously,  by an electron loop and a $\tau$ loop; i.e., the Feynman diagrams in Fig.~3. This contribution is given by the Feynman parametric integral:
\be
a_{\mu}^{\rm\tiny v.p.}(\text{\small Fig.~3}) = 
\frac{\alpha}{\pi}\int_{0}^{1} dx (1-x)\ 2\ 
\left[-e^2\Pi_{\mbox{\rm\footnotesize
R}}^{(e)}\left(\frac{-x^2}{1-x}m_{\mu}^2 \right) \right]\left[-e^2\Pi_{\mbox{\rm\footnotesize
R}}^{(\tau)}\left(\frac{-x^2}{1-x}m_{\mu}^2 \right) \right]\,,
\ee
where
\begin{equation}
(-ie)^2\Pi_{\mbox{\rm\footnotesize
R}}^{(l)}\left(\frac{-x^2}{1-x}m_{\mu}^2 \right)=	-\frac{\alpha}{\pi}\int_0^1 dz_l\ 2z_l\ (1-z_l)\ \ln\left(1+\frac{x^2}{1-x}\frac{m_{\mu}^2}{m_{l}^2}\ z_l \ (1-z_l)\right)\,,
\end{equation}
is a convenient one--Feynman parametric representation of the  on--shell renormalized  photon self--energy generated by a lepton $l$, at the one--loop level.

%%%%%%%%%%%%%%%%%%%%%%%%%%%%%%%%%%%%%%
\begin{figure}[h]

\begin{center}
\includegraphics[width=0.6\textwidth]{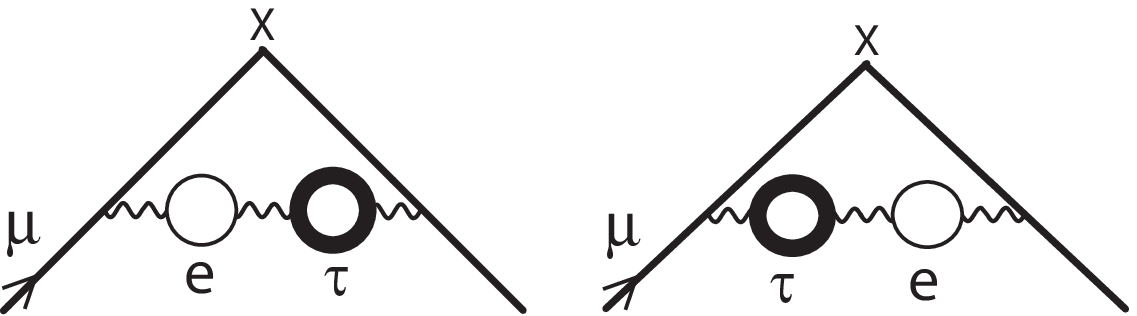}

\vspace*{0.25cm}
{\bf Fig.~3} {\it\small  Vacuum Polarization contributions to the Muon Anomaly\\ 
induced, simultaneously, by an electron--loop and a $\tau$--loop.
}
\end{center}

\end{figure}
%%%%%%%%%%%%%%%%%%%%%%%%%%%%%%%%%%%%%%
We have here a slightly  more complicated problem, with three Feynman parameters $x$,  $z_e$ and $z_{\tau}$ and two $\rho$--like parameters: 
$
\frac{m_{\mu}^2}{m_{e}^2}\gg 1\,,$ and 
$\frac{m_{\mu}^2}{m_{\tau}^2}\ll 1\,.$
In this case, it is better to apply another type of Mellin--Barnes representation to each log--factor in each photon self--energy as follows (see e.g. ref~\cite{Erde54}):
\begin{equation}
\ln\left(1+\frac{m_{\mu}^2}{m_{l}^2}X_{l}\right)=\frac{1}{2\pi i}\int\limits_{c -i\infty}^{c +i\infty} ds \left(\frac{m_{\mu}^2}{m_l^2} \right)^{-s}\left[X_{l}\right]^{-s} \frac{\pi}{s \sin(\pi s)}\,,
\end{equation}
where, here, $c=\text{Re}(s)\;\in\;]-1,0[$ defines the {\it fundamental strip} of the corresponding integration path along the imaginary axis.
This is a very useful Mellin--Barnes representation which also factorizes the dependence on the mass--ratio in the log--factor and which, to our knowledge, has not been used before in Quantum Field Theory calculations. The resulting Feynman parametric integrals can then be done trivially. This way, 
one finds that $a_{\mu}^{\rm\tiny v.p.}(\text{\small Fig.~3})$ has a double Mellin--Barnes integral representation: 
\begin{equation}
	a_{\mu}^{\rm\tiny v.p.}=\left(\frac{\alpha}{\pi}\right)^3 \ \frac{1}{2\pi i}\int\limits_{c_s-i\infty}^{c_s+i\infty} ds\  \left(\frac{m_{\mu}^2}{m_e^2}\right) ^{-s}
	\frac{1}{2\pi i}\int\limits_{c_t-i\infty}^{c_t+i\infty} dt\  \left(\frac{m_{\mu}^2}{m_{\tau}^2}\right)^{-t}\cM[a_{\mu}^{\rm\tiny v.p.} ](s,t)\,,
\end{equation}
with
\be\lbl{singsvpvp}
\cM[a_{\mu}^{\rm\tiny v.p.} ](s,t) 
=  8\ \frac{\Gamma(s)\Gamma(1-s)}{s}\frac{\Gamma(t)\Gamma(1-t)}{t} \frac{\left[\Gamma(2-s)\Gamma(2-t)\right]^2}{\Gamma(4-2s)\Gamma(4-2t)}
 \frac{\Gamma(1-2s-2t)\Gamma(2+s+t)}{\Gamma(3-s-t)}\,.
\ee

Here we are interested in the simultaneous singularities at the right of the {\it fundamental strip}~ $c_s=\text{Re}(s)\;\in\;]-1,0[$ (which encode the $m_{\mu}^2\gg m_e^2$ expansion)  and  at the left of the {\it fundamental strip}~ $c_t=\text{Re}(t)\;\in\;]-1,0[$ (which encode the $m_{\mu}^2\ll m_{\tau}^2$ expansion).
One can then proceed as follows. First we evaluate the singular series in the $t$--variable, for fixed $s$, and apply the {\it converse mapping theorem} to that series with the result
\begin{equation}
	\hspace*{-0.3cm}a_{\mu}^{\rm\tiny v.p.}\!\!\! \underset{ m_{\mu}\ll m_{\tau}}{\thicksim}\!\! \left(\frac{\alpha}{\pi}\right)^3 \! \frac{1}{2\pi i}\!\!\!\int\limits_{c_s-i\infty}^{c_s+i\infty}\!\!\! ds\! 
\left[\!\left(\frac{m_{\mu}^2}{m_e^2}\right)^{-s}\!\sum_{n\geq 1}\cA_{n}(s)\!\left(\frac{m_{\mu}^2}{m_{\tau}^2}\right)^{\! n}\!\!\! + \!
\left(\frac{m_{\tau}^2}{m_e^2}\right)^{-s}\!
\sum_{n\geq 0}\cB_{n}(s)\!\left(\frac{m_{\mu}^2}{m_{\tau}^2}\right)^{\! n+2}\!\right]\!\,,	
\end{equation}
where the $\left(\frac{m_{\tau}^2}{m_e^2}\right)^{-s}$ factor in the  $\cB$--series has been induced by the singular expansion of the  $\Gamma(2+s+t)$ factor in Eq.~\rf{singsvpvp} which generates the shift:
\begin{equation}
\left(\frac{m_{\mu}^2}{m_e^2}\right)^{-s}\left(\frac{m_{\mu}^2}{m_{\tau}^2}\right)^{-t}\Ra
\left(\frac{m_{\mu}^2}{m_e^2}\right)^{-s}\left(\frac{m_{\mu}^2}{m_{\tau}^2}\right)^{s+n+2}	=
\left(\frac{m_{\tau}^2}{m_e^2}\right)^{-s}
\left(\frac{m_{\mu}^2}{m_{\tau}^2}\right)^{n+2}\,;
\end{equation}
and
\begin{equation}\lbl{singetau}
\left(
\begin{array}{c}
\cA_{n}(s)\\ ~ \\ \cB_{n}(s)
\end{array}\right)
=8\ \frac{\Gamma(s)\Gamma(1-s)\Gamma(2-s)^2}{s\ \Gamma(4-2s)}
\left(
\begin{array}{l}
\frac{(-1)^{n+1}}{n} \frac{\Gamma(2+n)^2}{\Gamma(4+2n)}\frac{\Gamma(1-2s+2n)\Gamma(2+s-n)}{\Gamma(3-s+n)} \\
~ \\ 
\frac{(-1)^{n+1}}{n!} \frac{\Gamma(-2-n-s)\Gamma(3+n+s)\Gamma(4+n+s)^2\Gamma(5+2n)}{(s+n+2)\Gamma(8+2n+2s)\Gamma(5+n)} 
\end{array}\right)\,.
\end{equation}

The problem has now been reduced to  one of the type already discussed in the previous sections. In fact, even in this more complex case, we have been able to find the exact form of the full asymptotic expansion~\cite{Fr,Gr}. In practice, however, one is only interested in a few terms in that expansion. Applying once more the {\it converse mapping theorem}, one can immediately see that the singular behaviour:
\begin{equation}
\cA_{1}(s) \ \underset{s\ra 0}{{\thicksim}}\  \frac{2}{135}\frac{1}{s^2}+\frac{1}{135}\frac{1}{s}\,,
\end{equation}
controls the leading terms of $\cO\left(\frac{m_{\mu}^2}{m_{\tau}^2}\right)$. Terms of $\cO\left(\frac{m_{\mu}^2}{m_{\tau}^2}\right)^2$ have their source in the
singular behaviours:
{\setl
\bea
\cA_{2}(s) & \ \underset{s\ra 0}{{\thicksim}}\  & -\frac{1}{210}\frac{1}{s^3}+\frac{1}{504}\frac{1}{s^2}
	-\left(\frac{\pi^2}{630}-\frac{47}{30240}\right)\frac{1}{s}\,,
\\	\cB_{0}(s) & \ \underset{s\ra 0}{{\thicksim}}\  & \frac{1}{210}\frac{1}{s^3}-\frac{37}{22050}\frac{1}{s^2}
+\frac{39379}{2315250}\frac{1}{s}\,.
\eea}

\noi
Terms of $\cO\left(\frac{m_{\mu}^2}{m_{\tau}^2}\right)^3$ have their source in the singular behaviours:
{\setl
\bea
	\cA_{3}(s) & \ \underset{s\ra 0}{{\thicksim}}\  &  -\frac{4}{945}\frac{1}{s^3}-\frac{1}{4725}\frac{1}{s^2}
	-\left(\frac{4\pi^2}{2835}-\frac{11}{94500}\right)\frac{1}{s}\,, \\
\cB_{1}(s) & \ \underset{s\ra 0}{{\thicksim}}\  & \frac{4}{945}\frac{1}{s^3}-\frac{199}{297675}\frac{1}{s^2}
+\frac{2735573}{187535250}\frac{1}{s}\,;	
\eea}

\noi
and so on.
The next singularity at the right of the {\it fundamental strip} of  $\cA_{1}(s)$ is at $s\ra 1$:
\begin{equation}
\cA_{1}(s)\ \underset{s\ra 1}{{\thicksim}}\  -\frac{2}{15}\frac{1}{s-1}\,,
\end{equation}
which generates a term of $\cO\left[\left(\frac{m_{e}^2}{m_{\mu}^2}\right)\left(\frac{m_{\mu}^2}{m_{\tau}^2}\right)\right]$. Because of the factor $\Gamma(1-2s+2n)$ in Eq.~\rf{singetau}, the next--to--next singularity of  $\cA_{1}(s)$ at the right of the {\it fundamental strip} is at $s\ra 3/2$:
\begin{equation}
\cA_{1}(s)\ \underset{s\ra 3/2}{{\thicksim}}\  \frac{4\pi^2}{45}\frac{1}{s-\frac{3}{2}}\,,
\end{equation}
which generates a term of $\cO\left[\left(\frac{m_{e}^2}{m_{\mu}^2}\right)^{3/2}\left(\frac{m_{\mu}^2}{m_{\tau}^2}\right)\right]$; and so on. 

>From these results one can reconstruct, in a straightforward way, the asymptotic behaviour:
{\setl
\bea
	a_{\mu}^{\rm\tiny v.p.} & \underset{m_e \ll m_{\mu}\ll m_{\tau}}{\thicksim} & \left(\frac{\alpha}{\pi}\right)^3 \left\{ 
\frac{m_{\mu}^2}{m_{\tau}^2}	\left[\frac{2}{135}\ln\frac{m_{\mu}^2}{m_e^2}-\frac{1}{135}\right]
+\frac{2}{15}\frac{m_e^2}{m_{\tau}^2}-\frac{4\pi^2}{45}\frac{m_e^3}{m_{\tau}^2 m_{\mu}}
 \right. \nn \\
& & \hspace*{-2.5cm}+ \left(\frac{m_{\mu}^2}{m_{\tau}^2}\right)^2\left[-\frac{1}{420}\ln\frac{m_{\tau}^2}{m_{\mu}^2}\ln\frac{m_{\tau}^2\ m_{\mu}^2}{m_e^4}-\frac{37}{22050}\ln\frac{m_{\tau}^2}{m_e^2} +\frac{1}{504}\ln\frac{m_{\mu}^2}{m_e^2}+\frac{\pi^2}{630}-\frac{229213}{12348000}\right] \nn \\
& & \hspace*{-2.5cm} + \left(\frac{m_{\mu}^2}{m_{\tau}^2}\right)^3\left[-\frac{2}{945}\ln\frac{m_{\tau}^2}{m_{\mu}^2}\ln\frac{m_{\tau}^2\ m_{\mu}^2}{m_e^4}-\frac{199}{297675}\ln\frac{m_{\tau}^2}{m_e^2} -\frac{1}{4725}\ln\frac{m_{\mu}^2}{m_e^2}+\frac{4\pi^2}{2835}-\frac{1102961}{75014100}\right] \nn \\
& & \hspace*{-2.5cm} + \left(\frac{m_{\mu}^2}{m_{\tau}^2}\right)^4\left[-\frac{1}{594}\ln\frac{m_{\tau}^2}{m_{\mu}^2}\ln\frac{m_{\tau}^2\ m_{\mu}^2}{m_e^4}-\frac{391}{2058210}\ln\frac{m_{\tau}^2}{m_e^2} -\frac{19}{31185}\ln\frac{m_{\mu}^2}{m_e^2}+\frac{\pi^2}{891}-\frac{161030983}{14263395300}\right] \nn \\
 & & +\left. 
 \cO\left[\left(\frac{m_{\mu}^2}{m_{\tau}^2}\right)^5 \ln\frac{m_{\tau}^2}{m_{\mu}^2}\ln\frac{m_{\tau}^2\ m_{\mu}^2}{m_e^4} \right]+\cO\left(
\frac{m_e^2}{m_{\tau}^2} \frac{m_{\mu}^2}{m_{\tau}^2}\right)
 \right\}\,.
\eea}
\hspace*{-0.2cm} The first three lines are in agreement with the result obtained in ref.~\cite{CS98} using the {\it method of regions} ~\cite{BS98}. The terms in the fourth line are new; we have incorporated them just to show how easily higher order terms can be obtained, if necessary. By contrast,  the {\it method of regions} requires in this case the separate consideration of {\it five} different integration regions in the virtual loop--momenta with their appropriate Taylor expansions of propagators, plus the evaluation of a large number of integrals. That complexity is to be compared with the simplicity of the approach reported here.  

\vspace*{1cm}

{\bf Acknowledgments}

\vspace*{0.25cm}

\noi
We are grateful to Marc Knecht and Santi Peris for discussions and suggestions.
This work has been supported in part by TMR, EC-Contract No. HPRN-CT-2002-00311 (EURIDICE).

\end{document}